\documentclass[12pt]{article}

\usepackage{epsf}

\usepackage{graphicx}%

\begin{document}

\begin{center}
{\bfseries ADIABATIC CHEMICAL FREEZE-OUT AND WIDE RESONANCE MODIFICATION IN A THERMAL MEDIUM}

\vskip 5mm

K.A. Bugaev$^{1 \dag}$, D.R. Oliinychenko$^{1,2}$, E.G. Nikonov$^{3}$,  A.S. Sorin$^{2}$ and G.M. Zinovjev$^1$

\vskip 5mm

{\small
(1) {\it
Bogolyubov Institute for Theoretical Physics, Metrologichna str. 14$^B$, Kiev 03680, Ukraine
}
\\
(2) {\it
Bogoliubov Laboratory of Theoretical Physics, JINR, Joliot-Curie str.  6,   Dubna, Russia
}
\\
(3) {\it
Laboratory for Information Technologies, JINR, Joliot-Curie str. 6, 141980 Dubna, Russia
}
\\
$\dag$ {\it
E-mail: Bugaev@th.physik.uni-frankfurt.de
}}
\end{center}


\begin{abstract}
Here we develop  a  model equation of state which successfully  parameterizes  the 
thermodynamic functions of hadron resonance gas model at chemical freeze-out  and which allows us to naturally explain 
the   adiabatic chemical freeze-out  criterion. The present  model enables  us  to clearly  demonstrate  that at chemical freeze-out   the resulting hadronic  mass spectrum  used in the hadron resonance  gas model  is not an exponential-like, but a power-like.
We argue that such a property of  hadronic mass  spectrum at chemical freeze-out can be  explained by the two new effects found here  for 
wide resonances existing  in a thermal environment: the near threshold thermal resonance enhancement   and  the near threshold resonance sharpening. The effect of  resonance sharpening is studied for a sigma meson and our analysis shows that for the temperatures well below 92 MeV the effective width of sigma meson  is about 50 to 70 MeV. Thus, the effect of  resonance sharpening justifies  the usage of the sigma-like field-theoretical models for the strongly interacting matter equation of state at  such  temperatures. Also we argue  that the most optimistic hope to find  the quark gluon bags  experimentally  may be related to their sharpening and enhancement  in  a thermal medium. In this case  the  wide quark gluon bags  may appear directly or in decays as narrow  resonances that  are absent in the tables of elementary particles  and   that have the width about 50-150 MeV  and  the  mass about or above 2.5 GeV.
\end{abstract}

\vskip 3mm

{\bf I.} The hadron resonance gas model \cite{KABAndronic:05} is a reliable theoretical tool to extract information about the chemical freeze-out (FO) stage of the relativistic heavy ion collisions. However,  the question about the reliable chemical FO criterion has a long history \cite{KABAndronic:05,KABCleymansFO,KABBlaschke:11}. 
Very recently this question was thoroughly investigated again \cite{KABOliinychenko:12}, using the most sophisticated version of the hadron resonance gas model. Similarly to \cite{KABAndronic:12} it was found that none of the formerly suggested  chemical FO criteria \cite{KABCleymansFO} is robust, if the realistic particle table with the hadron masses up to 2.5 GeV is used.  However, in \cite{KABOliinychenko:12} the  criterion of   the adiabatic  chemical FO was suggested. In \cite{KABOliinychenko:12} it was also shown that despite an essential difference with the model used in \cite{KABAndronic:05}  the same conclusion on the constant entropy per particle at chemical FO 
is well  reproduced by  the chemical FO  parameters found in  \cite{KABAndronic:05,KABAndronic:12}.  
Furthermore, such a criterion of   chemical FO is also consistent with the best description 
of  the Strangeness Horn puzzle found recently in \cite{KABugaev:12}. 
Thus, it turns out that the criterion of the constant entropy per particle at chemical FO is, indeed, the reliable one. 
It is interesting that the constant entropy per particle at chemical FO was also found in \cite{KABTiwari:12},
but the way of hard core repulsion  used in this work is too different from the traditional  one used in the hadron resonance gas model \cite{KABAndronic:05, KABOliinychenko:12, KABAndronic:12} and,  hence, in contrast to the results of  \cite{KABOliinychenko:12,KABAndronic:12}, the model used in \cite{KABTiwari:12} leads to the simultaneous fulfillment of a few chemical FO criteria. 

Despite the long history of this question until recently there was  no a single attempt  to understand what is the physical reason behind any of the chemical FO criterion. Only last year this  problem  got  an  adequate interest of  theoreticians \cite{KABBlaschke:11}.
Here we  develop an alternative approach and present a simple model equation of state which not only well  describes the chemical FO thermodynamic parameters and reproduces the constant value of the entropy per particle at the chemical FO, 
but in addition it allows us to elucidate the real mass spectrum of mesons and baryons that generate such a criterion. 
As it will be shown below the real mass spectrum of  hadrons, i.e. the density of hadronic states, is very much different from the Hagedorn mass spectrum which  is traditionally expected to emerge already  for  hadrons  with masses above 1.2 GeV \cite{KABHagedorn1,KABHagedorn2,KABHagedorn3}.  However, below we  demonstrate that the real mass spectrum of hadrons
extracted from the adiabatic  chemical FO model  is not an exponential, but a power-like and the reason for such a behavior is an  existence of many wide resonances. In this work   we also show how at chemical FO  a  thermal medium    essentially  modifies 
the resonance mass distribution in case of large width leading  to   their  narrowing near the threshold.  
Based on these findings we suggest that the  quark-gluon bags may be observed at the NICA energy range  as the narrow resonances (width about 50-150 MeV) with the mass 
about or above 2.5  which are absent in the tables of elementary particle properties.

\vspace*{0.22cm}

{\bf 2.} It is, indeed, a remarkable fact that at the chemical FO the entropy per particle deviates from 7.18 by about 8  \% while the  center of mass energy of collision changes from $\sqrt{s_{NN}} =2.2 $ GeV to $\sqrt{s_{NN}} =7$ TeV \cite{KABOliinychenko:12}.  In order to find  the validity reason of   the adiabatic chemical FO criterion  
 we parameterize the mesonic (with subscript $M$) and baryonic (with subscript $B$)  pressure as follows 
\begin{eqnarray}\label{EqI}
&&p_M= C_M ~ T^{A_M} ~ \exp\left[ \frac{\mu_M- m_M}{T}\right] \,,\\
\label{EqII}
&&p_B= C_B ~\, T^{A_B}\, ~ \exp\left[ \frac{\mu_B- M_B}{T}\right] \,,
\end{eqnarray}
assuming that the integrated mass spectrum of mesons and baryons can be represented by the constants
$C_a$, $A_a$, $m_a$, and $\mu_M$ with $a \in \{M, B \}$.
Here $T$ denotes the chemical FO temperature and $\mu_B$ is the corresponding baryonic chemical potential.
 The pressure of antibaryons is, evidently, related to that one of baryons as $p_{\bar B}(T, \mu_B) = p_{B}(T, - \mu_B)$.
Note  that Eqs.  (\ref{EqI}) and (\ref{EqII}) represent the mixture of gases of  massive particles with the temperature dependent number of degrees of freedom without the hard-core repulsion. 
The thorough inspection \cite{KABAndronic:05,KABOliinychenko:12} shows that for the realistic values of hadronic hard-core radii below 0.45 fm  the effect of hard-core repulsion can be safely neglected for temperatures below 150-170 MeV. This fact  is 
implemented in Eqs. (\ref{EqI}) and (\ref{EqII}). 

Using the standard thermodynamic identities from Eqs. (\ref{EqI}) and (\ref{EqII})  one can find   the particle densities  $\rho_a = \frac{\partial p_a}{\partial \mu_a} = p_a/T $ with $a \in \{M, B, \bar B \}$ and entropy density $s  \equiv \frac{\partial (p_M+p_B + p_{\bar B})}{\partial  T}$ at chemical FO and get the following expression for entropy per particle 
\begin{eqnarray}\label{EqIII}
&&\frac{s}{\rho_{P}} = \frac{ (A_M +\frac{m_M - \mu_M}{T} )\,\rho_M~ + ~ (A_B +\frac{M_B}{T}) (\rho_B +
 \rho_{\bar B})~ -  ~\frac{\mu_B}{T}(\rho_B - \rho_{\bar B}) }{\rho_M +   \rho_B + \rho_{\bar B}}    \,,
\end{eqnarray}
where  the particle density is denoted as $\rho_P \equiv \rho_M +   \rho_B + \rho_{\bar B}$. In order to qualitatively explain the reason of constant ratio ${s}/{\rho_{P}}$ we note that there exist two regions
where  the baryonic chemical potential roughly linearly decreases with the FO temperature \cite{KABOliinychenko:12,KABAndronic:05}, i.e. $\mu_B(T) \simeq \mu_B (T_0) + \mu^\prime_B  (T-T_0 ) \simeq \mu_B (T_0) - \mu^\prime_B\, T_0 + \mu^\prime_B\,T \simeq \mu_B^0  + \mu^\prime_B \,T$.  Taking the latter into account, one can write (\ref{EqIII}) as
\begin{eqnarray}\label{EqIV}
&&\frac{s}{\rho_{P}} \simeq \frac{ (A_M +\frac{m_M - \mu_M}{T} )\,\rho_M~ + ~ (A_B - \mu^\prime_B  +\frac{M_B- \mu_B^0}{T}) (\rho_B +
 \rho_{\bar B})~  +  ~2 \, \frac{\mu_B}{T}\rho_{\bar B} }{\rho_M +   \rho_B + \rho_{\bar B}}    \,.
\end{eqnarray}
Consider, first, the chemical FO temperatures below 135 MeV \cite{KABAndronic:05, KABOliinychenko:12}. 
Note that in (\ref{EqIV}) the term $\frac{\mu_B}{T}\frac{\rho_{\bar B}}{\rho_M +   \rho_B + \rho_{\bar B}} \ll 1  $ since at large values of the baryonic chemical potentials the antibaryons are absent, while their number becomes comparable with the number of baryons for $\frac{\mu_B}{T} \rightarrow 0$. Now it is clear that one can get rid of the 
strong temperature dependence in  (\ref{EqIV}), if  the following conditions are obeyed
\begin{eqnarray}\label{EqV}
&& \frac{m_M - \mu_M}{T}  \simeq 0\, , \quad 
\frac{M_B- \mu_B^0}{T}   \simeq 0  \,.
\end{eqnarray}
Then the entropy per particle at chemical FO can be cast as 
\begin{eqnarray}\label{EqVI}
&&\frac{s}{\rho_{P}} \simeq A_B - \mu^\prime_B - ( A_B - \mu^\prime_B - A_M)  \frac{\rho_M}{\rho_M +   \rho_B + \rho_{\bar B}}   \simeq  A_B - \mu^\prime_B \, ,
\end{eqnarray}
i.e. up to a small correction the entropy per particle is constant. Strictly speaking, the above rough estimates should be valid only   for low chemical FO temperatures, when the number of mesons is, indeed, a correction to the number of baryons.  However, at high chemical FO temperatures  the baryonic chemical potential  is also linearly dependent on $T$ \cite{KABOliinychenko:12,KABAndronic:05}, although with another value of the derivative $\tilde \mu^\prime_B$. Then at high chemical FO temperatures the entropy density per particle  (\ref{EqV}) becomes $\frac{s}{\rho_{P}} \simeq A_B - \tilde \mu^\prime_B - ( A_B - \tilde \mu^\prime_B - A_M)  \frac{\rho_M}{\rho_M +   \rho_B + \rho_{\bar B}}   \simeq  A_M  + ( A_B - \tilde \mu^\prime_B - A_M)  \frac{(\rho_B + \rho_{\bar B})}{\rho_M +   \rho_B + \rho_{\bar B}}  \simeq Const$, since at these temperatures the fraction of baryons and antibaryons is almost a constant. 
Thus, the entropy per particle at chemical FO is  basically defined by  the power $A_a$ of the dominated degrees of freedom ($a =B$  for low T and $a =M$ for high T)  and by the  speed of baryonic chemical potential change with the temperature change, i.e. by $\mu^\prime_B$.

\begin{figure}[h]
 \centerline{
 \includegraphics[width=50mm,height=50mm]{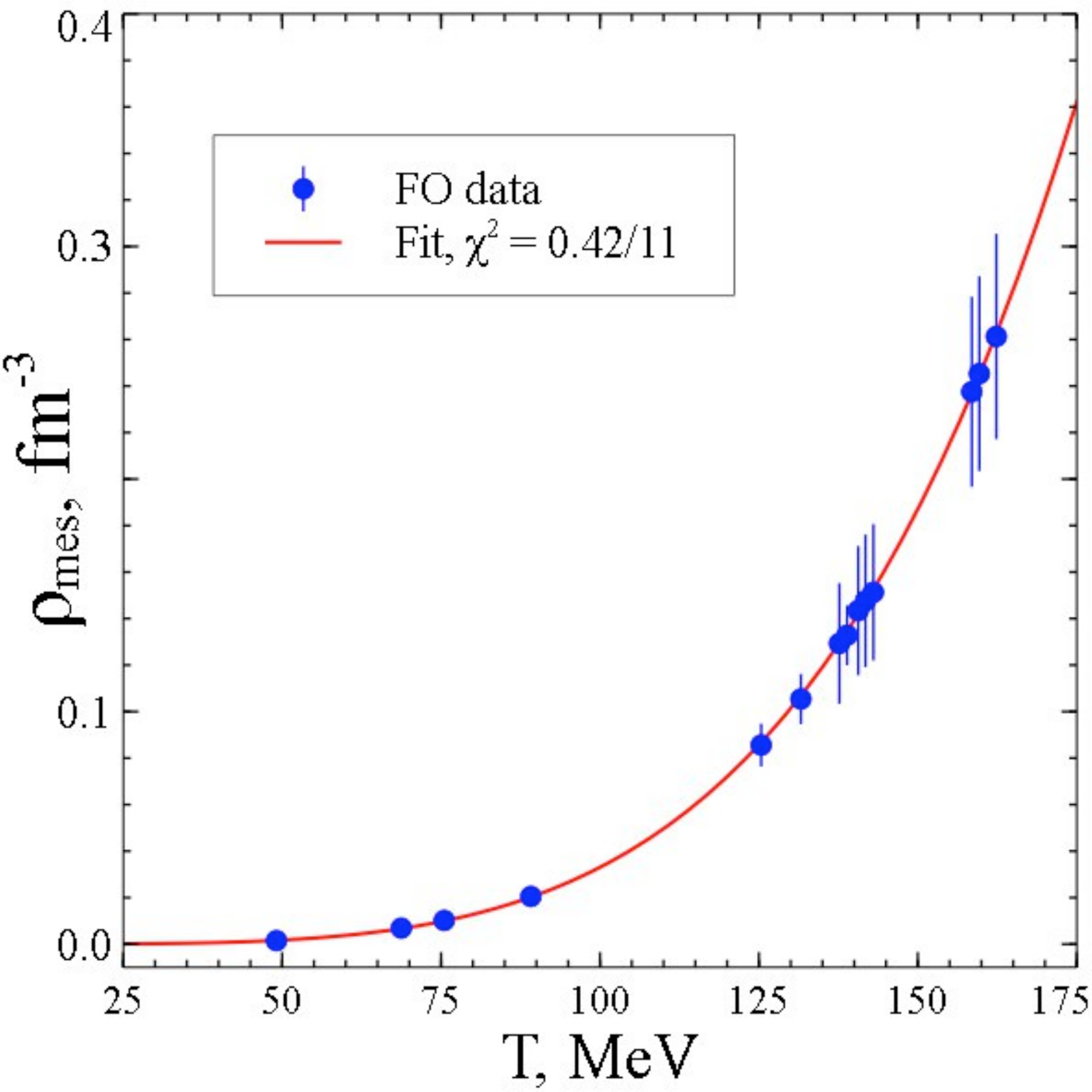} \hspace*{1.cm}
 \includegraphics[width=50mm,height=50mm]{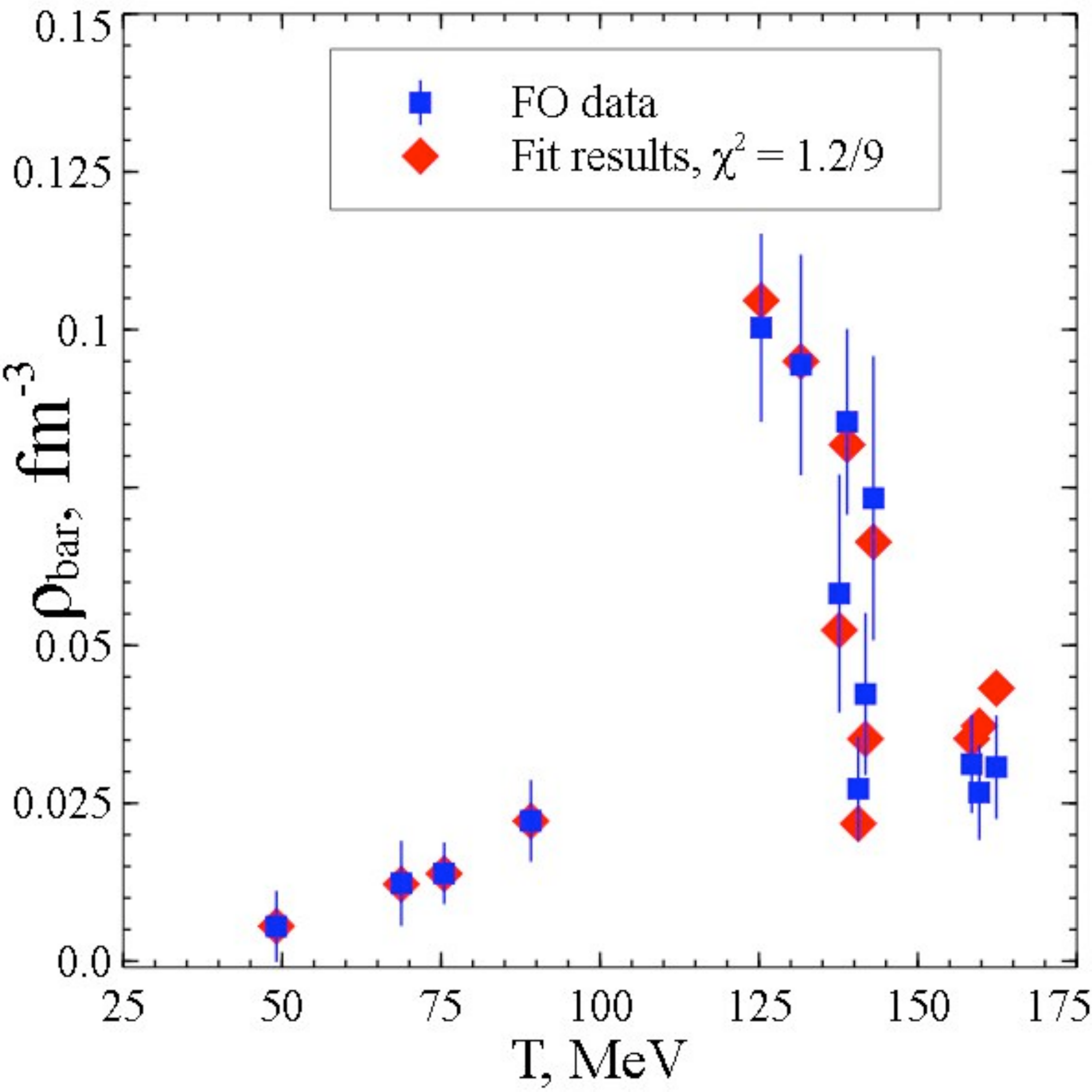}
 }
 \caption{Temperature dependence of  meson particle density $\rho_M$ (left panel) and baryon and antibaryon  particle density $(\rho_B + \rho_{\bar B})$ (right panel) for the developed model. 
 The data points with error bars  are the results of  the fit obtained in
 \cite{KABOliinychenko:12} for  the experimental hadronic multiplicities.}
 \label{KABfig1}
\end{figure}

The above simplified consideration is well supported by a quantitative analysis. To demonstrate this in Fig. 
\ref{KABfig1} we show the results 
of fit of  the particle densities obtained in \cite{KABOliinychenko:12} for   chemical FO. As one can see from Fig. \ref{KABfig1}  the quality of the chemical FO data for mesons (left panel) is much higher than that one 
of the baryons and antibaryons and, hence, the corresponding description of meson densities is excellent, while the description of baryon-antibaryon densities is good since  the value of  $\chi^2/dof =7.8/11$ is  acceptable. The major part of this $\chi^2/dof $ value is generated by the
three data points of  the baryon-antibaryon density  for the highest temperatures.
We found the same minimum, if  these three points are excluded from the fit, although 
in this case
the  corresponding value of  $\chi^2/dof$ is much better $\chi^2/dof = 1.2/8$. 
Our analysis shows that such a defect of the present model 
can be repaired by  the excluded volume corrections.
However, 
 here we prefer to keep the model as simple as possible in order to demonstrate the main idea and, hence, hereafter  we do not consider the last three data points into further fitting. 

The results of the density fit are  as follows 
\begin{eqnarray}\label{EqVII}
&&M_B  \simeq 800.5 \pm 30 \, {\rm MeV}\,,  \quad   \quad m_M-\mu_M  \simeq - 5 \pm 5 \, {\rm MeV}
\,, \quad  \\
\label{EqVIII}
&& C_B \simeq (2.624 \pm 0.191) \cdot 10^{-9}  \, \frac{{\rm \small MeV}^{1-A_B}}{{ \rm fm}^3} \,, \quad  C_M \simeq (7.61 \pm 0.12) \cdot 10^{-9}  \, \frac{{\rm \small MeV}^{1-A_M}}{{ \rm fm}^3}  \,, \\
\label{EqIX}
&&A_B \simeq 6.097 \pm 0.38\, ,  \quad   A_M \simeq 5.31 \pm 0.14\, . 
\end{eqnarray}
Note that the conditions (\ref{EqV}) are well reproduced by  the fitting of the densities. It may, however, look strange that 
for the  vanishing value of  mesonic chemical potential  the meson mass parameter $m_M$ in (\ref{EqVII}) can be negative. 
We note that the results of fitting show that condition (\ref{EqV}) for mesons is numerically well established,  since  the minimal value of the chemical FO temperature is about 50 MeV. Therefore, the result (\ref{EqVII})  for mesons means that within the error bars one has 
$\mu_M = 0$ and $m_M = 0$. 

\begin{figure}[h]
 \centerline{
 \includegraphics[width=50mm,height=50mm]{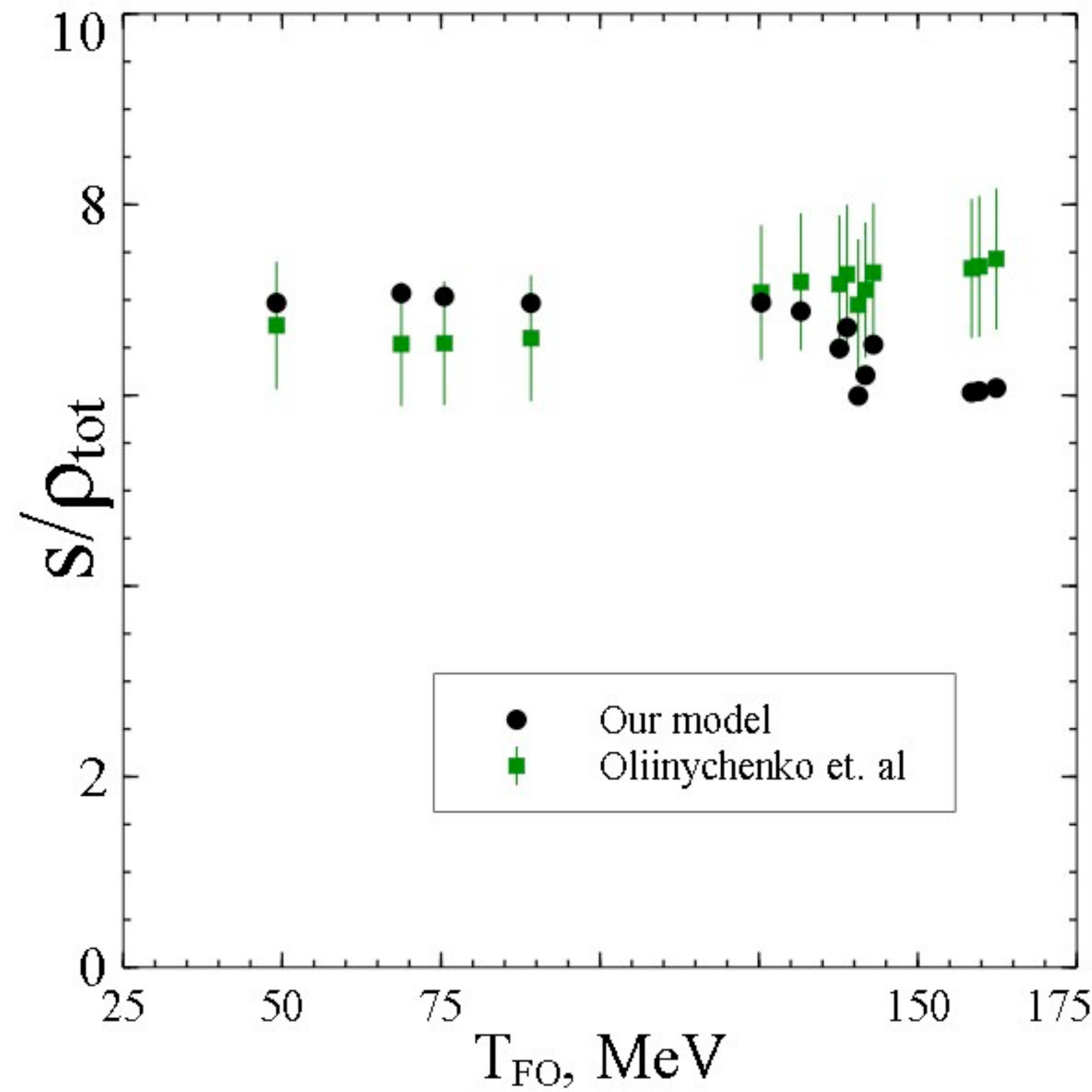} \hspace*{1.cm}
  \includegraphics[width=50mm,height=50mm]{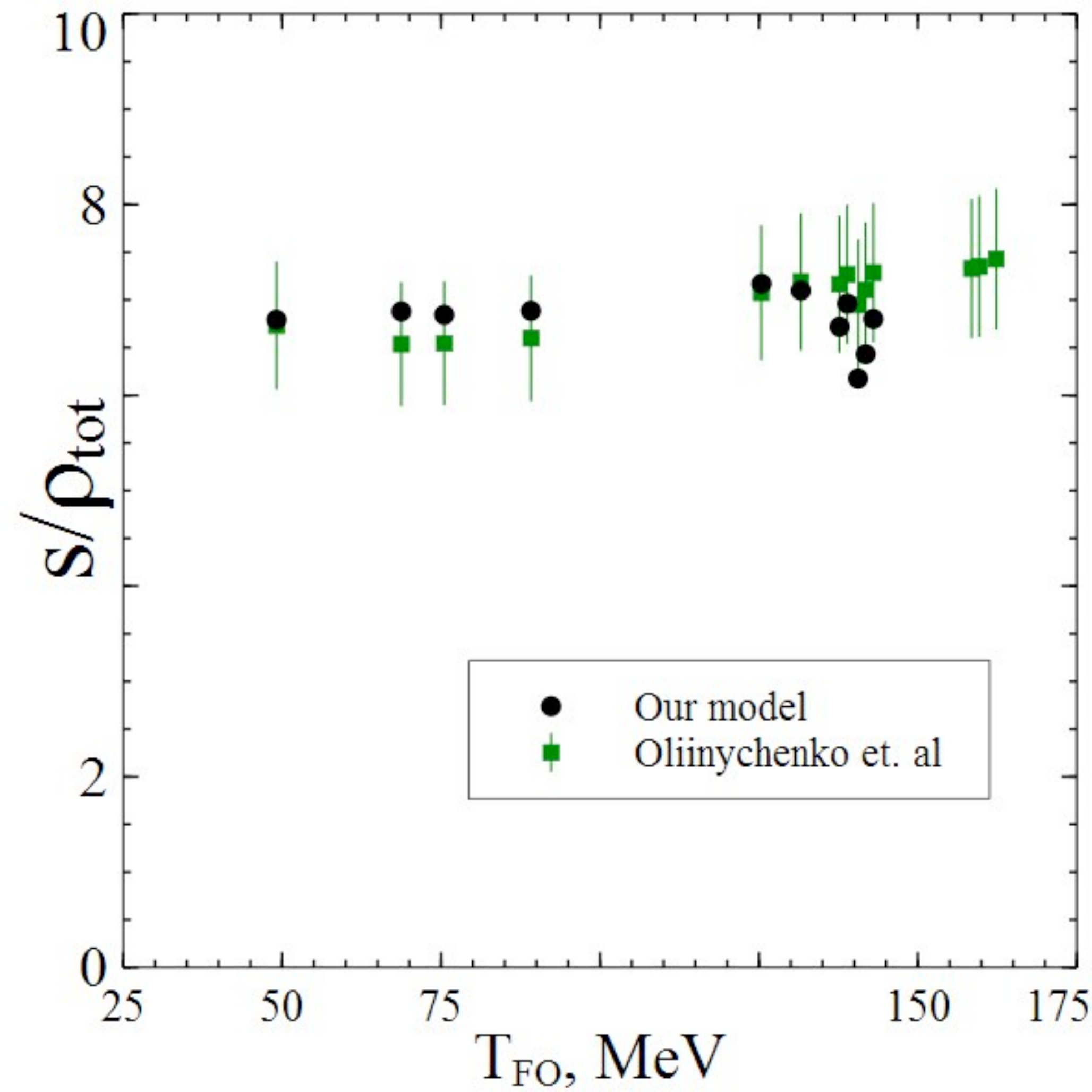}
 }
 \caption{Left panel: Comparison of the chemical FO  temperature dependence of  the entropy per particle ${s}/{\rho_P}$  for the developed model (circles) and  for the chemical FO data found in \cite{KABOliinychenko:12}
 (squares with error bars).  
 Note that this ratio was just calculated using the parameters of Eqs. (\ref{EqVII}) - (\ref{EqIX}), i.e.  no fitting was  used. Right panel: the same ratio as in the left panel, but fitted together with particle densities.
}
 \label{KABfig2}
\end{figure}

Using the above  parameters we calculated (no additional fit!) the entropy density and found its ratio to the particle density. 
The results are shown in the left panel of Fig. \ref{KABfig2}. From this figure one can see that only the three points at highest chemical FO temperatures which were not included into a fit  are not well reproduced. 
Of course, as it is seen  from the right panel of Fig. \ref{KABfig2} the additional fitting of  the ratio $s/\rho_P$  can improve the quality description of all analyzed points  and one gets  $\chi^2/dof \simeq 0.47$. 

{\bf 3.} The model equations (\ref{EqI})  and  (\ref{EqII}) allow us to directly find the mass spectrum of hadrons. 
The question of  whether the experimental mass spectrum of hadrons given in the Particle Data Group tables  coincides with  the spectrum suggested by R. Hagedorn   is of great interest nowadays \cite{KABHagedorn1,KABHagedorn2,KABHagedorn3}. However, almost all  discussions of the hadron mass spectrum simply ignore the width of  resonances, whereas 
the large resonance width may essentially modify the spectrum \cite{KABStoecker:86,KABStashek:87,KABStashek:90,KABHeinz:91,KABDirk:91,
KABBlaschke:04,KABBlaschke:05,FWM:08} and it may be even  responsible for an absence of  heavy excited resonances 
in the empirical  mass spectrum of hadrons \cite{FWM:08, FWM:09}. Therefore, it is interesting to 
study the effective mass spectrum of hadrons having the real width. Since in the   hadron resonance gas model \cite{KABAndronic:05, KABOliinychenko:12}   the resonance width is considered explicitly, its effect is  also implicitly accounted  for in the model equations (\ref{EqI}) and (\ref{EqII}).

In order to elucidate the effective mass spectrum of baryons we 
 rewrite their  density  in terms of the mass integral of the momentum integral  of the Boltzmann distribution function.  Since the effective baryon mass $M_B $ in (\ref{EqVII}) is large compared to 
the maximal value of the chemical FO temperature, one can safely use the non-relativistic approximation for the momentum integration. This means that the factor $T^\frac{3}{2}$ should be assigned to the momentum integration, while the remaining T-dependence of the baryon (and antibaryon) particle  density can be identically cast as
\begin{eqnarray}
\rho_B &=& C_B \, T^{A_B-1}\,  \exp\left[ \frac{ \mu_B- M_B }{T} \right]
 =
 C_B\, \exp \left[ \frac{\mu_B}{T} \right]  \int\limits_{M_B}^\infty d m \, \frac{(m-M_B)^{A_B-3.5} }{ \Gamma(A_B-2.5) } \,T^\frac{3}{2}  \, \exp\left[ -\frac{m}{T}\right]  
 \nonumber  \\
\label{EqX}
& \simeq &   (2\pi)^\frac{3}{2} C_B
 \, \exp \left[  \frac{\mu_B}{T}  \right]  \int\limits_{M_B}^\infty d m \, 
 \frac{ (m-M_B)^{A_B-3.5} }{m^\frac{3}{2} \, \Gamma(A_B-2.5) } 
 \int \frac{d^3 k}{ (2 \pi)^3 }   \exp \left[ -\frac{ \sqrt{k^2 + m^2} }{T} \right]  \,,
\end{eqnarray}
where in the last step of derivation we accounted for the fact that besides $\exp \left[ -\frac{m}{T}\right]$ the momentum  integration in non-relativistic case generates the factor $ (2 \, \pi \, m \,T)^\frac{3}{2}$. Here $\Gamma(A) $ is the usual gamma function and, hence, one has $\Gamma(3.6) \simeq 3.717 $.   

Since the original mass integral in (\ref{EqX}) is the Laplace integral, then  the representation (\ref{EqX}) uniquely  defines  the $T$-dependence of  baryonic density and vice versa.  Thus, from (\ref{EqX})
one concludes that at chemical FO the effective mass spectrum of baryons and antibaryons   is rather 
a power-like than an exponential-like:
\begin{eqnarray}\label{EqXI}
\frac{\partial   \varrho_B (m) }{\partial m} \simeq \frac{(2\pi)^\frac{3}{2}\, \hbar^3 \, \, C_B}{\Gamma(3.6)} \, \frac{(m-M_B)^{2.6}}{  m^{\frac{3}{2}} } \,, \quad \Rightarrow \quad  \varrho_B (m) \biggl|_{m \gg M_B} \sim m^{2.1} \,. 
\end{eqnarray}

\begin{figure}[htbp]
\centerline{\includegraphics[height=7 cm]{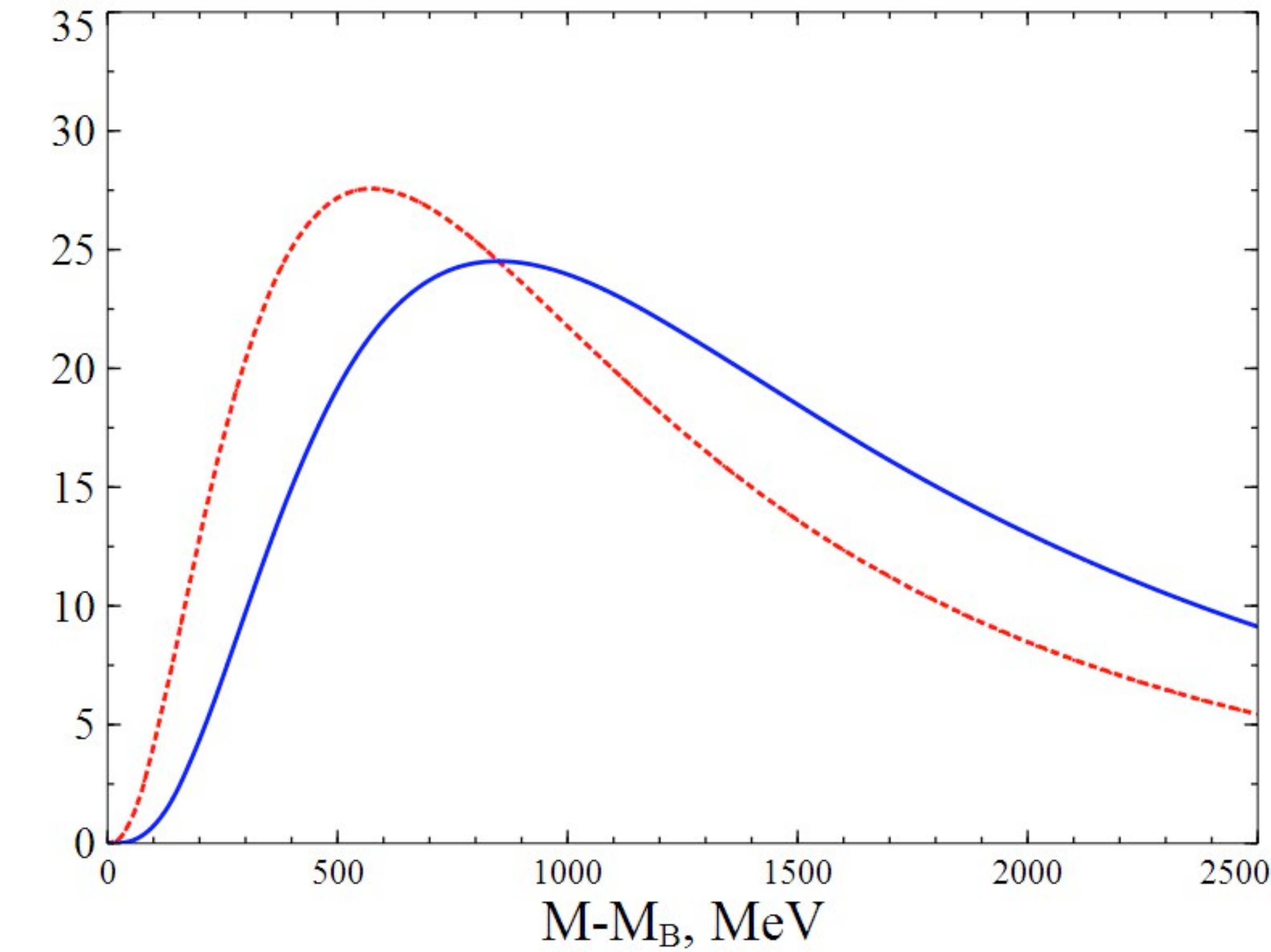}   
}
 \caption{Ratio of  the baryonic mass spectrum $\varrho_B (m)$  of  the present model found at chemical FO defined by   (\ref{EqXI}) to the mass spectrum of all hadrons $ \varrho_B^{exp} (m) $ taken from 
the Particle Data Group  \cite{KABPDG:08} and   parameterized according to  (\ref{EqXII}) \cite{KABHagedorn4}.
The full curve shows the ratio of integrated spectra $\varrho_B (m)/\varrho_B^{exp} (m)$, while the dotted curve demonstrates the ratio of  the corresponding densities of states. The mass   $M_B = 800.5$ MeV is taken from Eq. (\ref{EqVII}).
}
 \label{KABfig3}
\end{figure}

{\bf 4.} Comparing the baryonic  mass spectrum $\varrho_B (m) \equiv \int\limits_{M_B}^m  d z  \, 
\frac{\partial   \varrho_B (z) }{\partial z}$ obtained from the  density of  states  (\ref{EqXI}) with the parameterization  
\begin{eqnarray}\label{EqXII}
 \varrho_B^{exp} (m)  \simeq  \left[ \frac{m}{537\, \rm MeV}\right]^{5.72} \,, 
\end{eqnarray}
suggested in \cite{KABHagedorn4} to describe the experimental hadronic mass spectrum, one finds a great disagreement between them! Indeed, as one can see from  Fig. \ref{KABfig3} these two mass spectra may deviate
from each other by  
about 25 times at $m \simeq 1650$ MeV and their asymptotic behaviors are  completely different,
since $ \varrho_B (m)  \simeq m^{2.1}  $ in (\ref{EqXI}), while $ \varrho_B^{exp} (m)  \simeq m^{5.72}$
in (\ref{EqXII}).  Naturally, there arise two questions, ``Does it  mean that one of these two mass spectra is wrong?" and ``What is the reason for  so huge difference between  these mass spectra?"
In order to demonstrate that none of these two mass spectra is wrong and that at chemical FO the large  difference between these two  hadronic mass spectra   is due to the width of hadronic resonances, we  consider the Gaussian mass attenuation  instead of the Breit-Wigner one that is used in the  actual  simulations since in this case  the evaluation is more transparent. 
Also  such a treatment would allow us to obtain some important conclusions on the mass spectrum of 
quark-gluon  (QG) bags which according to \cite{FWM:08,FWM:09} should unavoidably have the Gaussian mass attenuation. 
Note also   that  the estimates below provide us with the lower limit,
since  the Gaussian  mass distribution  vanishes  much faster than the Breit-Wigner one. The typical term of  the $k$-resonance that enters into  the mass spectrum of the baryonic  particle density  of the hadron resonance gas model   is given by  $ F_k (\sigma_k) \exp{ \left[ \frac{\mu_B}{T} \right] }$ \cite{KABAndronic:05,KABOliinychenko:12},  where 
\begin{eqnarray}
\label{EqXIII}
%
F_k (\sigma_k) & \equiv &  g_k \int\limits_{0}^\infty  d m \,  \frac{\Theta\left(  m - M_k^{Th} \right) }{N_k (M_k^{Th})} 
 \, \exp \left[ - \frac{(m_k -m)^2}{2\, \sigma_k^2}  \right] 
 \int \frac{d^3 p}{ (2 \pi \hbar)^3 }   \exp \left[ -\frac{ \sqrt{p^2 + m^2} }{T} \right]  \,. \quad \quad 
\end{eqnarray}
Here $m_k$ is the mean mass of the $k$-th resonance, $g_k$ is its degeneracy factor, $\sigma_k$ is the Gaussian width which defines the true width of  such  a resonance as $\Gamma_k = Q\,  \sigma_k$ (with $Q \equiv 2 \sqrt{2\, \ln2}$) and the normalization factor is defined via the threshold mass $M_k^{Th}$   of the dominant channel as $N_k (M_k^{Th}) \equiv  \int\limits_{M_k^{Th}}^\infty  d m \, 
 \, \exp \left[ - \frac{(m_k -m)^2}{2\, \sigma_k^2}  \right] $.
For the narrow resonances the term $F_k (\sigma_k)$ converts into the usual thermal density of particles, i.e. for $\sigma_k \rightarrow 0$ one has $F_k \rightarrow  g_k \, \phi(m_k, T)$, where the following notation is used $ \phi(m,  T) \equiv  \int \frac{d^3 p}{ (2 \pi \hbar)^3 } \exp \left[ -\frac{ \sqrt{p^2 + m^2} }{T} \right] $.  

A usage of Eq. (\ref{EqXIII})  was heavily criticized in \cite{KABFriman}, but we find such a critique
absolutely inadequate for the states below the chemical FO.  First of  all, we note that  in the approach of \cite{KABFriman} the effect of  medium  cannot be switched off at any finite particle density or temperature. 
This means that according to  the treatment of  \cite{KABFriman} (and many similar works!) all the hadrons whose momentum spectra are frozen due to the absence of any strong interaction between them should keep their momentum dependent width and mass which  they acquired at the moment of kinetic FO up to they are captured by the  detectors. Hence, according to \cite{KABFriman} all hadrons measured by detectors, including the stable ones, are some resonances that `feel' a thermal medium in which they were produced long after the medium is gone. Second, all the `effects' which the authors of \cite{KABFriman} claim to be of principal  physical importance are reduced to a slight (by about 20 MeV) shift of the mass attenuation peak and small change of  its shape for the $\Delta_{33}$ resonance, although the corresponding modification of the nucleon and pion properties the authors of \cite{KABFriman} simply ignore.  In our opinion such modifications of the mass attenuation of   the $\Delta_{33}$ resonance compared to the `crude' approximation of  Eq.  
(\ref{EqXIII})  cannot be measured  in heavy ion experiments even for such narrow resonances as $\Delta_{33}$. Therefore, a serious discussion of  similar `effects'  for heavy hadronic resonances whose mass and width are  often known  with the accuracy of 100 MeV  or 200 MeV (or worse) \cite{KABPDG:08} does not make any sense. Thus, at the present state of  art 
there is no alternative to a physically transparent Eq. (\ref{EqXIII}) to be used at and after the moment of 
chemical FO. 

\begin{figure}[h]
\centerline{\includegraphics[width=7 cm]{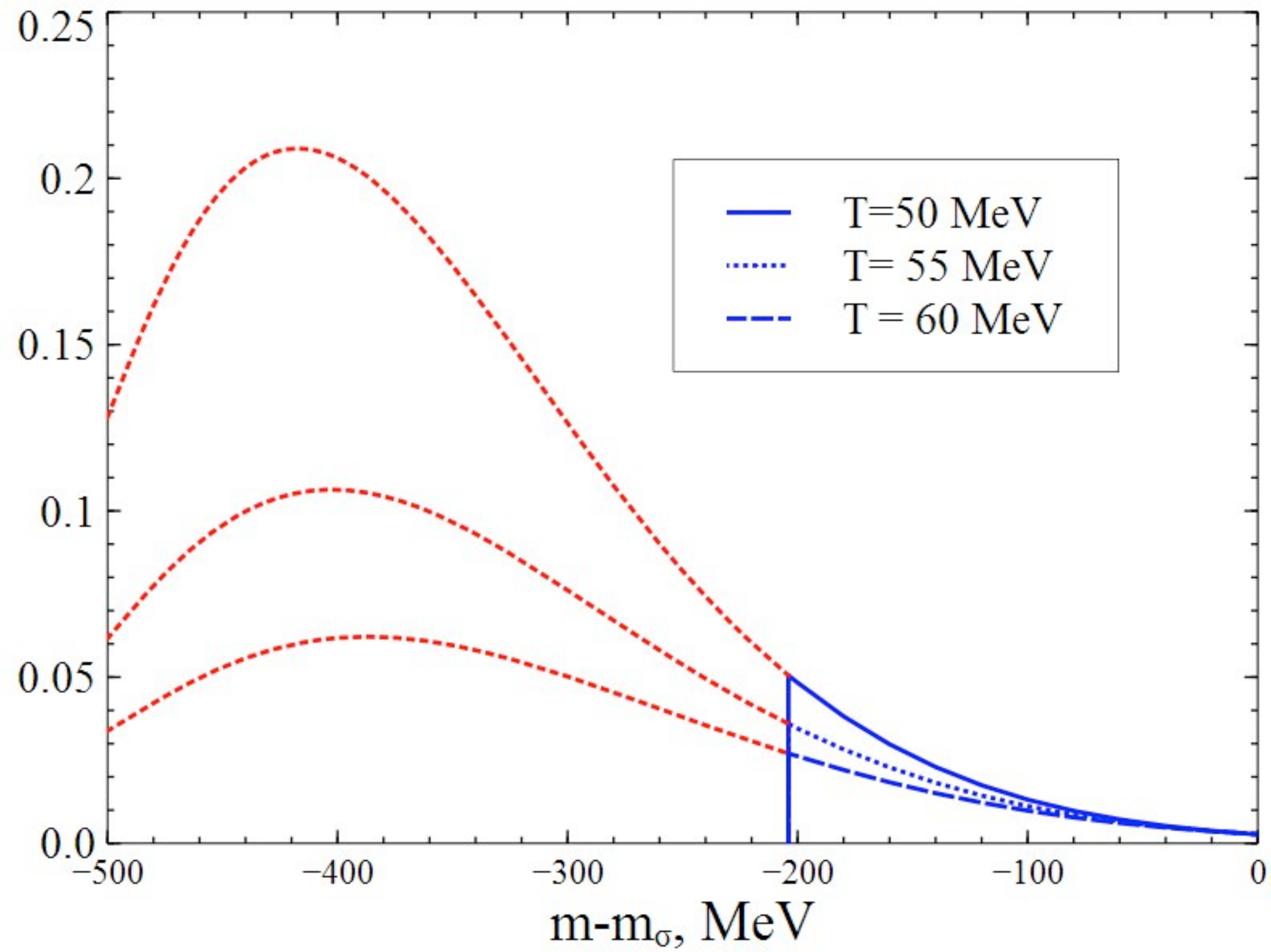}   
\includegraphics[width=7 cm]{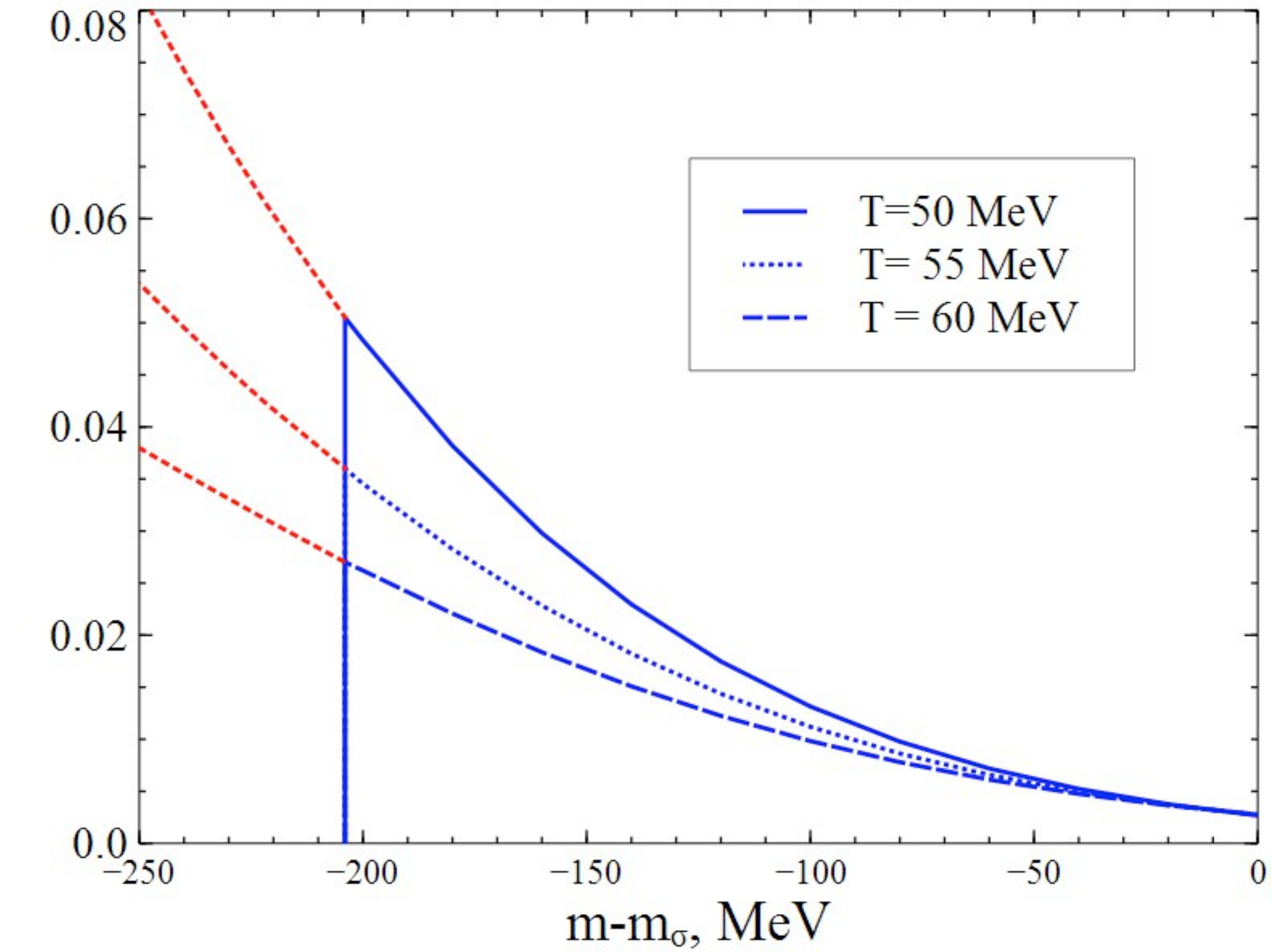} 
}
 \caption{Temperature dependence of  the mass distribution  $f_k(m)/ \phi(m_\sigma, T)  $ (in units of $1/$MeV, see Eq. (\ref{EqXIV})) for $\sigma$-meson with the mass $m_\sigma = 484$ MeV, the width $\Gamma_\sigma = 510$ MeV  \cite{KABSigma:07} and $M_\sigma^{Th} = 2 \, m_\pi \simeq 280$ MeV.  In the left panel the short dashed curves below the two pion threshold (vertical line at $m-m_\sigma = -204$ MeV)  show the mass attenuation  which does not contribute into the particle density (\ref{EqXIV}). From the right panel one can see the effect of  wide resonance sharpening near the threshold, i.e. an appearance of a narrow peak in the resulting mass distribution on the right hand side of the threshold which resembles an  icy slide. For different temperatures this  mass attenuation   is shown by the solid, the short dashed 
 and the long dashed curves.  The $\sigma$-meson  effective width   was found numerically from these mass attenuations: 
 $\Gamma^{eff}_\sigma (T=50\, {\rm MeV}) \simeq 62.5$ MeV, 
 $\Gamma^{eff}_\sigma (T=55\, {\rm MeV}) \simeq 71.5$ MeV and 
  $\Gamma^{eff}_\sigma (T=60\, {\rm MeV}) \simeq 82.5$ MeV. 
 }
 \label{KABfig4}
\end{figure}

The momentum integral in (\ref{EqXIII}) can be written using the non-relativistic approximation $\phi(m, T) \simeq
\left[ \frac{m\, T}{2\, \pi \, \hbar^2} \right]^\frac{3}{2}\exp \left[   -\frac{  m } {T} \right] $. Then  to simplify the mass integration of  (\ref{EqXIII})  one can make the full square in it from the  powers of  $(m_k -m)$ and get  
\begin{eqnarray}
\label{EqXIV}
\hspace*{-0.5cm}
F_k (\sigma_k) & \equiv & g_k  \int\limits_{0}^\infty  d m \, f_k ( m)  \simeq   \tilde g_k    \int\limits_{0}^\infty  d m \,  \frac{\Theta\left(  m - M_k^{Th} \right)}{N_k (M_k^{Th})}
 \, \exp \left[ - \frac{(\tilde m_k -m)^2}{2\, \sigma_k^2}  \right]  
 \left[ \frac{m\, T}{2 \pi   \hbar^2} \right]^\frac{3}{2}\exp \left[   -\frac{  m_k } {T} \right]  \,, ~
\end{eqnarray}
where the following notations for an effective resonance  degeneracy $\tilde g_k$   and  for an effective resonance mass $\tilde m_k$
\begin{eqnarray}\label{EqXIVa}
\tilde g_k     & \equiv &  g_k  \exp \left[  \frac{\sigma_k^2}{2 \, T^2} \right] = g_k  \exp \left[  \frac{\Gamma_k^2}{2 \,Q^2\, T^2} \right] \\
 \tilde m_k  & \equiv & m_k - \frac{ \sigma_k^2 }{T} = m_k - \frac{ \Gamma_k^2 }{Q^2\, T} 
\label{EqXIVb} 
\end{eqnarray}
are used.  From Eq.  (\ref{EqXIV})
one can  see that  the presence of the width, firstly,  may  strongly  modify the degeneracy  factor $g_k$ and,
 secondly,  it may essentially shift the maximum of the mass attenuation towards  the threshold or even below it.
 There are two corresponding effects which we named as {\it the near threshold thermal resonance enhancement} and  {\it the near threshold resonance sharpening}.   
 These effects formally appear due to the same reason as the  famous Gamow window for the thermonuclear reactions in stars \cite{KABGamow1, KABGamow2}:
 just above the resonance decay threshold 
 the integrand in (\ref{EqXIV}) is a product of two functions of a virtual resonance mass $m$, namely, the Gaussian attenuation is an increasing 
 function of $m$, while the Boltzmann exponent strongly decreases above the threshold. The resulting attenuation of their product has a maximum,
 whose shape, in contrast to the usual Gamow window,  may be extremely asymmetric due to the presence of the threshold.  Indeed, 
 as one can see from Fig. \ref{KABfig4} the resulting mass attenuation of a  resonance may acquire  the form of  the sharp and narrow peak that is 
 closely resembling
 an icy slide. 
 Below we discuss   these two effects in some details. 
Qualitatively the same effects appear, if in (\ref{EqXIV})  one uses the Breit-Wigner 
 resonance mass attenuation instead of the Gaussian one. 

From the  definitions of  the effective resonance mass (\ref{EqXIVb}) and the effective resonance degeneracy (\ref{EqXIVa})  one can see that the effects of their change are strong for $T \ll \sigma_k$. This can be clearly seen from  Fig. \ref{KABfig4}, which demonstrates both of the above effects at low temperatures  for  the $\sigma$-meson. A simple analysis shows that the effect of resonance sharpening is strongest, if the threshold mass is shifted to the convex part of the Gaussian distribution in (\ref{EqXIV}), i.e. for $M_k^{Th} \ge \tilde m_k$ or for the temperatures  $T$ below $T^+_k \equiv \frac{\sigma_k^2}{m_k - M_k^{Th}} \equiv \frac{\sigma_k}{\beta_k}$. 
To  demonstrate the effect of the width sharpening we  list 
a few typical examples for baryons  in the Table 1.   For $T < T^+_k$ and for $m > M_k^{Th}$ the Gaussian  mass distribution in (\ref{EqXIV}) can be safely approximated as $\exp \left[ - \frac{(\tilde m_k -m)^2}{2\, \sigma_k^2}  \right]  \approx \exp \left[ - \frac{(\tilde m_k - M_k^{Th})^2}{2\, \sigma_k^2} \, - \, \frac{(\tilde m_k - M_k^{Th})}{ \sigma_k^2}   (m - M_k^{Th}) \right]$. Now recalling the standard 
definition of the width for the function $f (x) = \Theta(x)\, Const \, \exp\left[ - b \, x  \right]$, one obtains  
the temperature dependent  resonance effective width near the threshold as 
\begin{eqnarray}
\label{EqXV}
\Gamma_k^N (T)  & \simeq  &  \frac{\ln(2)}{ \frac{1}{T} - \frac{\beta_k}{\sigma_k} } \equiv   \frac{\ln(2)}{ \frac{1}{T} -  \frac{1}{T^+_k} }    \,, 
\end{eqnarray}
 since for such a distribution function $f(x)$ one gets $f(\ln(2)/b) = f(0)/2$. Note that in evaluating (\ref{EqXV}) we neglected the additional $m^{1.5}$-dependence in (\ref{EqXIV}), but one can readily check that numerically such a correction is small.  The rightmost column in  Table 1  
 demonstrates that  Eq. (\ref{EqXV}), indeed, provides an accurate estimates for $T < T_k^+$.  The results of  Table 1   also justify the usage of $\sigma$-meson and the field-theoretical models based on the well known $\sigma$-model for temperatures well  below $T^+_\sigma \simeq 92$ MeV.  Of course, the present approach which is developed for the chemical FO stage, when the inelastic reactions except for resonance decays are ceased to exist, cannot be applied for earlier stages of heavy ion  collisions. However, here we would like to stress  that an inclusion of the large width of $\sigma$-meson in the field-theoretical models of the strongly interacting matter equation of state is very necessary. From the above analysis one can see that the large width inclusion can generate some new important physical effects like the wide resonance sharpening in a thermal medium. 

\begin{table}[h]
\bigskip
\begin{tabular}{|c|c|c|c|c|c|c|c|c|c|c|c|c|}
\hline
     Hadron &  $m_k$  & $\Gamma_k$ &   Decay    &   $M_k^{Th}$  & $\beta_k$ & $T_k^+$   &  
     $\Gamma_k^{eff}$  & $\Gamma_k^{N}$ \\ 
                   &  (MeV)       &            (MeV)          &   ~ channel ~    &        (MeV)      &  &        (MeV)  &       (MeV) &             (MeV) \\\hline
         $\sigma$-meson    &         484        &     510    &      $\sigma \rightarrow \pi \pi$   &    280  & 0.942 & 91.9 & 
         62.5 &  67.3 \\  \hline
          $P_{33}$     &   1232  &    120   &    $\Delta \rightarrow \pi N$    &    1080  & 2.98 & 11.6 &  43.5 & N/A  \\  \hline
          $P_{11}$     &   1440  &    350   &    $N \rightarrow \pi N$    &    1080  & 2.42 & 38.74 &  129.5 & N/A  \\  \hline
          $P_{33}$     &   1600  &    350   &    $\Delta \rightarrow \pi \Delta$    &    1372  & 1.53 & 50.4  & 68.7 & 80.8 \\  \hline
          $P_{33}$     &   1600  &    350   &    $\Delta \rightarrow \pi N$    &    1080  & 3.5 & 30.3 &  280. & N/A \\  \hline
            $G_{17}$     &   2190  &    500   &    $\Delta \rightarrow \rho N$    &    1710  & 2.26 & 57.8  & 74.6 & 81.8 \\  \hline
\end{tabular}
\vspace*{0.3cm}
\caption{
The parameters of several hadronic resonances together with their  decay  channels  that are used to determine the quantities $\beta_k$ and $T_k^+$. The last two columns show the corresponding effective width at temperature $T=50$ MeV found, respectively, numerically from Eq. (\ref{EqXIII}) and analytically  from Eq. (\ref{EqXV}), when it can be applied. 
}
\label{table1}
\end{table}

\begin{figure}[ht]
\centerline{\includegraphics[width=8 cm]{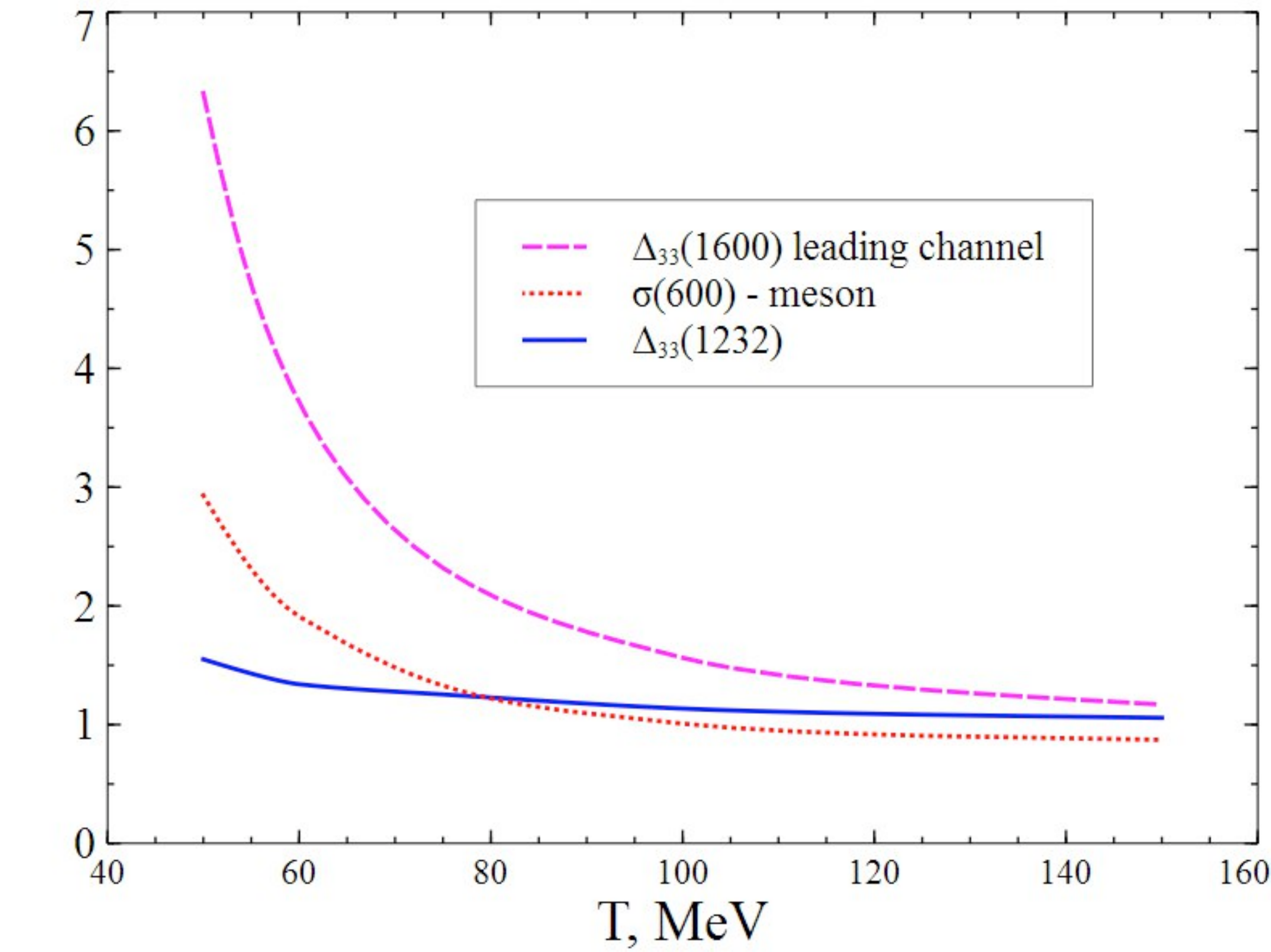}   
\includegraphics[width=8 cm]{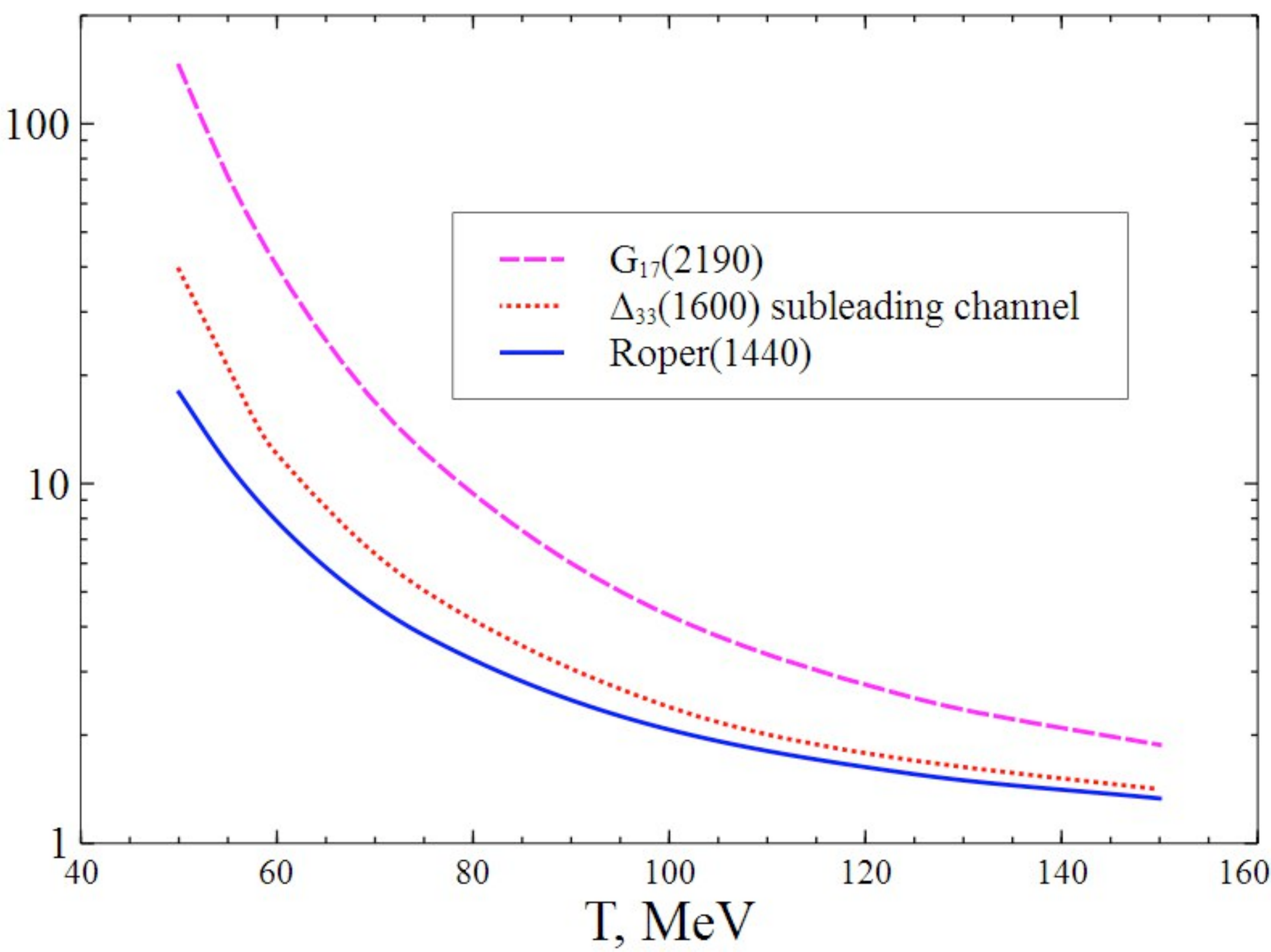} 
}
 \caption{Temperature dependence of  the resonance  enhancement. The ratio $R(T) = \frac{f_k}{\phi (m_k,T)}$  is shown for the hadronic resonance decays given in Table I.  For wide resonances the effect of enhancement can be huge. 
  }
 \label{Geff_ot_T}
\end{figure}

From  Fig. \ref{Geff_ot_T} one can see that the resonance enhancement  can be, indeed,  huge for wide ($\Gamma \ge  450$ MeV) and medium wide ($\Gamma \simeq 300-400$ MeV) resonances. 
This effect naturally explains the strong temperature dependence of  hadronic pressure 
(\ref{EqI}) 
at chemical FO, which in its turn generates the power-like  mass spectrum of  baryons (\ref{EqXI}).
Clearly, the same is true for the mesonic pressure  (\ref{EqII}) and  mesonic mass spectrum in a thermal environment. However, we believe that  a detailed study  of  such a phenomenon  requires a special investigation. 

\vskip 2.5mm

{\bf 5.} The first important conclusion from the analysis above is that there is no sense to discuss the mass spectrum of 
hadronic resonances, empirical or Hagedorn, without a treatment of  their width. Furthermore, the same is true 
for  the QG bags which, according to the finite width model \cite{FWM:08,FWM:09},  are heavy and wide  resonances 
with mass $M_B$ larger than $M_0 \simeq 2.5$ GeV and with  the mean width of the form $\Gamma_B \simeq \Gamma_0 (T) \left[  \frac{M_B}{M_0}  \right]^\frac{1}{2}$,  where  $\Gamma_0 (T)$ is a monotonically increasing function  of $T$ and  $\Gamma_0 (T=0) \in [400; 600]$ MeV.  
This range of   $\Gamma_0 (T=0)$  values corresponds to the cross-over temperatures $T_{co} \simeq 170-200$ MeV \cite{FWM:08,FWM:09} for vanishing baryonic density. The value $\Gamma_0 (T=0)=400$ MeV   is well consistent with the results of the 
present days lattice QCD thermodynamics \cite{KABKarsch,KABWupBuD:2009}, but there is no 
guaranty that the lattice QCD data will not change in the future. Therefore, below  we consider the whole range of 
values for the width $\Gamma_0 (T)$ analyzed in \cite{FWM:08,FWM:09}. 

There are two interesting features of  QG bags which are related to the above treatment. Thus, from the results of  \cite{FWM:08,FWM:09} and from (\ref{EqXV})  one can find   the  temperature $T^+_B$  for the QG bags  as 
\begin{eqnarray}
\label{EqXXX}
&T^+_B \simeq  \frac{ \Gamma_0^2 (T) }{Q^2\, M_0 \left(1 - \xi_B  \right) } \simeq 
\frac{1}{\left(1 - \xi_B  \right) }\, \cdot \, \left\{ 
\begin{array}{ll}
11.5-26.\,\, \rm MeV  \,,
&
{\rm if} ~ ~\Gamma_0 \simeq 0.4-0.6\,\, {\rm GeV} ~{\rm at}~ T=0 \,,  \\
46-104\,\, \rm MeV  \,,
&
{\rm if} ~ ~\Gamma_0 \simeq 0.8-1.2\,\, {\rm GeV} ~{\rm at}~ T=90  \,,  ~ \\
140-315\,\, \rm MeV  \,,
&
{\rm if} ~ ~\Gamma_0 \simeq 1.4-2.1\,\, {\rm GeV} ~{\rm at}~ T=170  \,,
\end{array}
\right. \label{EqXVIII}
\end{eqnarray}
where  $\xi_B \equiv \frac{M_B^{Th}}{M_B}$ denotes the ratio of the leading threshold mass  $M_B^{Th}$ of the bag to  its  mean mass  $M_B$. In (\ref{EqXVIII}) the temperature $T$ is given in MeV.  Clearly for different bags the range of  $\xi_B$ value can be  between 0 and 1. 
Therefore,  according to above results  the bags with $\xi_B \rightarrow  1$ should have been essentially enhanced and 
sharpened as the ordinary resonances. Moreover,  according to (\ref{EqXV}) in this case for $T  \ll T_B^+$ the QG bags should have had a small width $\Gamma_B^N \simeq \frac{T \, T_B^+}{T_B^+ - T} \ln(2)$ and, hence, such QG bags should have been stable or, in other words, these bags should have been observed! The reason why such bags are not observed in the experiments is naturally explained by the finite width model  \cite{FWM:08,FWM:09}:
it is due to the effect called as {\it the subthreshold suppression}, i.e. a huge  suppression (of about fifteen to sixteen orders of magnitude compared to light hadrons!) of the QG bags for temperatures below the  half of the traditional Hagedorn temperature $T_H$ (for more details see a discussion after Eq. (42) in \cite{FWM:09}).  Such a suppression is 
a manifestation of  the color confinement in terms of the QG bag width \cite{FWM:09}.

On the other hand Eq. (\ref{EqXXX}) also shows that the only hope to observe the QG bags exists, if $\xi_B \rightarrow 1$. Then for chemical FO temperatures much below $T^+_B$ such bags could have sufficiently long  eigen lifetime of about $\tau_B \sim \frac{1}{\Gamma_B^N} \simeq \frac{T_B^+ - T }{T \, T_B^+\, \ln(2)} \le  \frac{1}{T \,  \ln(2)} $.
Substituting  $T \simeq 0.5 \, T_H \in [80; 90]$ MeV in the last inequality and using the estimate of  Eq. (\ref{EqXVIII}) for $T=90$ MeV with $\xi_B = 0.9$, one finds the most optimistic estimates for the QG bag eigen lifetime as  $\tau_B \le 3.3 \pm 0.3 $ fm/c. 
These estimates allow us to make the second important conclusion  that {\it   an  appearance  of   sharp  resonances (baryonic or/and  mesonic) with the width 
in the interval between 50 to 70 MeV}  at the chemical FO  temperatures close to $T_{QGB} \simeq  85 \pm 5$ MeV  {\it that have  the mass 
above 2.5 GeV and that are absent in the tables of particle properties would be  a clear signal of the QG bag formation.}
Their possible appearance at chemical FO as metastable states of  finite systems created in relativistic nuclear collisions  is justified by the finite width model 
\cite{FWM:08,FWM:09}.
At higher temperatures such QG bags can be formed too, but their width is larger and lifetime is shorter. The limiting values of  $\xi_B$ at $T$ for which the effect of resonance sharpening can exist is determined by the relation $\Gamma_0^2 (T)/T \ge Q^2\, M_0 \left(1 - \xi_B  \right)$ from which one can see that the condition $\xi_B \rightarrow 0.9$ can be relaxed, but  in this case the temperature of chemical FO  gets higher.

In addition one  has to account for  the statistical probability of  the QG bags appearance at a given temperature $T$. Relatively to the nucleon  the statistical   probability of the QG bag of mass $M_B$ is about $W = \left[ \frac{M_B}{M_N} \right]^{1.5} \exp \left[ \frac{(M_N-M_B)}{T} \right] \, R_B(T)$, where $M_N \simeq 940$ MeV is the nucleon mass and  $R_B(T)$ is the resonance enhancement factor in a thermal medium.  For $T= 140$ MeV and $M_B = M_0 \simeq 2.5$ GeV one gets $W_B \simeq 3.85 \cdot 10^{-5} R_B$. Our analysis shows that for such temperatures  the typical  resonance effective width values  are about $\Gamma_B \simeq 100 -150$ MeV while the typical values of the resonance enhancement factor is about $ R_B \simeq 10-100$   and, hence, compared to nucleon  the relative statistical probability of such QG bags is about $W_B \simeq 3.85 \cdot (10^{-4}- 10^{-3})$, which is  essentially larger than the relative probability of the $J/\psi$ meson  $W_{J/\psi} \simeq 1.19 \cdot 10^{-6}$ at the same temperature. Note that the chemical FO temperature $T \simeq 140$ MeV corresponds to the highest SPS energy of collision at which the $J/\psi$ mesons are safely measured. 
Note that the  chemical FO temperatures about $T_{QGB} \simeq  80-140$ MeV correspond to the center of mass  energy of collision $\sqrt{s_{NN}} \in [4; \, 8]$ GeV \cite{KABAndronic:05,KABOliinychenko:12}, which is in the range of the Dubna Nuclotron and NICA  energies of collision.
This energy range sets the  most promising  kinematic limit for the QG bag searches. 
\vspace*{0.22cm}

{\bf 6.} In conclusion,  here we present  a simple model equation of state which successfully  parameterizes  the 
thermodynamic functions of hadron resonance gas model at chemical FO. Such a model with the temperature dependent number of hadronic degrees of  freedom  allowed us to naturally explain 
the   adiabatic chemical freeze-out  criterion 
that was  found previously. In addition, this model allowed us to find out that the effective mass spectrum of baryons used in the resonance hadron gas model  is not an exponential-like, but a power-like. Evidently, the same conclusion is valid for mesons, but such an analysis will be published elsewhere.  

In order to give a reason for the obtained   difference between the effective baryonic mass spectrum and the empirical hadronic mass spectrum we analyzed the behavior of  wide resonances in a thermal environment  and    found  two new effects occurring, if  the chemical FO  temperature is small  compared to the resonance width: {\it the near threshold thermal resonance enhancement} and  {\it the near threshold resonance sharpening}.
Further analysis showed that for the temperatures well below $92$ MeV the $\sigma$-meson can be rather narrow and it  has  the effective  width of about 50 to 70 MeV. Thus, accounting for the $\sigma$-meson large width in a thermal medium  allows us to justify the usage of the $\sigma$-like field-theoretical models for the strongly interacting matter equation of state for temperatures well below $92$ MeV.
Finally, we argued that the most optimistic hope to find  the QG bags  experimentally  would be related to their sharpening and enhancement  by  a thermal medium. Then the QG bags may appear directly or in decays as narrow  resonances of the width about 50-150 MeV that have  the  mass about or above 2.5 GeV and  that  are absent in the tables of elementary particles. 

\vspace*{3mm}

\noindent
{\bf Acknowledgments.} The authors appreciate the valuable comments of  D. B.  Blaschke. 
K.A.B., D.R.O. and G.M.Z. acknowledge  the partial  support of the Program `On Perspective Fundamental Research in High Energy and Nuclear Physics' launched by the Section of Nuclear Physics  of National Academy of  Sciences of Ukraine.
The work of  E.G.N. and A.S.S.  was supported in part by the Russian
Foundation for Basic Research, Grant No. 11-02-01538-a.


\end{document}